\documentclass[3p,times,procedia]{elsarticle}
\usepackage{nupha_ecrc}
\usepackage[utf8x]{inputenc}

\newcommand{\rt}{{\mathbf{r}}}
\newcommand{\xt}{{\mathbf{x}}}
\newcommand{\bt}{{\mathbf{b}}}

\newcommand{\yt}{{\mathbf{y}}}
\newcommand{\zt}{{\mathbf{z}}}

\newcommand{\tr}{\, \mathrm{Tr} \, }

\newcommand{\nc}{{N_\mathrm{c}}}

\newcommand{\gev}{\ \textrm{GeV}}

\newcommand{\lqcd}{\Lambda_{\mathrm{QCD}}}
\newcommand{\as}{\alpha_{\mathrm{s}}}

\newcommand{\xpom}{{x_\mathbb{P}}}

\newcommand{\der}{\mathrm{d}}

\newcommand{\A}{{\mathcal{A}}}

\volume{00}

\firstpage{1}

\journalname{Nuclear Physics A}

\runauth{}

\jid{nupha}

\jnltitlelogo{Nuclear Physics A}

\usepackage{amssymb}

\usepackage[figuresright]{rotating}

\begin{document}

\begin{frontmatter}



\dochead{XXVIIth International Conference on Ultrarelativistic Nucleus-Nucleus Collisions\\ (Quark Matter 2018)}

\title{Energy and system size dependence of subnucleonic fluctuations}


\author[jyu]{Heikki Mäntysaari}
\author[bnl]{Björn Schenke}
\address[jyu]{Department of Physics, P.O. Box 35, 40014 University of Jyv\"askyl\"a, Finland}
\address[bnl]{Physics Department, Brookhaven National Laboratory, Upton, NY 11973, USA}

\begin{abstract}
The energy evolution of the fluctuating proton structure is studied by solving the JIMWLK renormalization group equation. The initial condition at moderate $x$ is obtained by fitting the charm reduced cross section data from HERA, requiring that the proton size remains compatible with the diffractive vector meson production measurements. Additionally, we show that the nucleon shape fluctuations are visible in exclusive vector meson production off nuclei.
\end{abstract}




\end{frontmatter}


\section{Introduction}
The small-$x$ structure of protons and nuclei is most precisely studied in deep inelastic scattering (DIS) experiments, where the structure of the target is probed by scattering of electrons. In inclusive DIS at high energy, the total gluon density of the target is measured. To date, the most precise knowledge of the parton densities is provided by combined work of the HERA experiments H1 and ZEUS~\cite{Abramowicz:2015mha,H1:2018flt}.  

One possibility to obtain a more differential picture of the partonic structure is provided by exclusive scattering, where the total momentum transfer can be measured. As the momentum transfer is the Fourier conjugate to the impact parameter, these processes provide access to the geometric structure, and in case of incoherent processes, also to the event-by-event fluctuations~\cite{Miettinen:1978jb, Mantysaari:2016ykx, Mantysaari:2016jaz}.

When the target proton is replaced by a nucleus, it becomes possible to access event-by-event fluctuations at different distance scales. At long distances (probed at small momentum transfer in exclusive scattering), the overall density fluctuations, and the fluctuating positions of the protons and neutrons contribute. When the momentum transfer increases, the nucleon substructure becomes visible, and can have a significant contribution to the incoherent vector meson production cross section~\cite{Mantysaari:2017dwh}. 

Currently, the only high-energy DIS data is provided by HERA experiments with proton targets. Additionally, ultraperipheral heavy ion collisions especially at the LHC have made it possible to study photoproduction processes off protons and nuclei at high energies, but in limited kinematics. In the future, the Electron Ion Collider~\cite{Accardi:2012qut} will make precise measurements of proton and nuclear structure over a wide kinematical range, making it possible to study non-linear QCD effects and to constrain the initial state of heavy ion collisions. In this context, we have studied in Ref.~\cite{Mantysaari:2018zdd} the energy evolution of the small-$x$ gluonic structure of the proton, and in Ref.~\cite{Mantysaari:2017dwh} the effect of nucleon shape fluctuations within the nuclei.

\section{Structure functions and vector meson production}
A convenient way to describe deep inelastic scattering at high energies is provided by the Color Glass Condensate (CGC) effective theory of high-energy QCD~\cite{Gelis:2010nm}.  In the CGC, the DIS process is conveniently described in the dipole picture in which the incoming virtual photon splits into a quark-antiquark dipole, which subsequently scatters off the strong color field of the target. The total virtual photon-target cross section can be written as
\begin{equation}
\label{eq:gamma-xs}
	\sigma^{\gamma^* p}_{L,T} = 2 \sum_f \int \der^2 \bt \der^2 \rt \frac{\der z}{4\pi} \left|\Psi^f_{L,T}(r,z,Q^2)\right|^2 \langle N(\rt, \bt, x) \rangle,
\end{equation}
where $\Psi^f$ is the wave function describing the $\gamma^*\to q\bar q$ splitting, and $\langle N(\rt, \bt, x) \rangle$ is the average dipole-target scattering amplitude. In terms of Wilson lines, which contain all the information about the target, the dipole amplitude is written as $1- N(\xt, \yt, x) = \tr V(\xt)V^\dagger(\yt) / \nc$ with $\rt=\xt-\yt$ and $\bt=(\xt+\yt)/2$. The sum $f$ runs over quark flavors, $z$ is the longitudinal momentum fraction of the photon carried by the quark, and $T,L$ refer to the transverse and logitudinal photon polarization, respectively.

In case of exclusive vector meson production, one has to also model the formation of a vector meson from the quark-antiquark pair, which is parametrized in terms of the vector meson light cone wave function $\Psi_V$. The scattering amplitude for exclusive vector meson production reads (see e.g.~\cite{Kowalski:2006hc})
\begin{equation}
\label{eq:diff_amp}
 \A^{\gamma^* p \to V p}_{T,L}(\xpom,Q^2, {\bf \Delta})= 2i\int \der^2 \rt \int \der^2 \bt \int \frac{\der z}{4\pi}   (\Psi^*\Psi_V)_{T,L}(Q^2, \rt,z) e^{-i[\bt - (1-z)\rt]\cdot {\bf \Delta}}  N(\rt,\bt,\xpom),
\end{equation}
where $\Psi^*\Psi_V$ denotes the overlap between the virtual photon and the vector meson wave functions. In case of coherent scattering, in which the target remains intact, the cross section is $\der \sigma / \der t = |\langle  \A^{\gamma^* p \to V p}_{T,L} \rangle |^2/(16\pi)$. In the incoherent scattering the target breaks up, and the cross section becomes a variance: $\der \sigma / \der t = \left[ \langle |  \A^{\gamma^* p \to V p}_{T,L} |^2 \rangle -   |\langle  \A^{\gamma^* p \to V p}_{T,L} \rangle |^2 \right]/(16\pi)$. As a variance, it measures the amount of event-by-event fluctuations in the scattering amplitude (and consequently in the impact parameter profile of $N$), whereas the coherent cross section is sensitive to the average shape of the target.

The energy evolution of the Wilson lines can be obtained by solving the JIMLWK renormalization group equation. In the Langevin form, which is suitable for numerical calculations, it can be written as
$
\frac{\der}{\der y} V_\xt = V_\xt (i t^a) \left[
\int \der^2 \zt\,
\varepsilon_{\xt,\zt}^{ab,i} \; \xi_\zt(y)^b_i  + \sigma_\xt^a 
\right].
$
Here $\xi$ is a random Gaussian noise with coefficient $\varepsilon$, and $\sigma$ is a deterministic drift term. For detailed expressions and details about the implementation, the reader is referred to Ref.~\cite{Mantysaari:2018zdd} and references therein.

\section{Results}
The initial condition for the JIMWLK evolution of the proton is obtained by fitting the HERA combined measurement of the charm structure function data~\cite{H1:2018flt}. We choose to fit that dataset instead of the full inclusive structure function data, as we want to suppress the contributions from dipoles larger than the proton size which have large contribution to the structure function $F_2$ in case of light quarks even at high $Q^2$ (see~\cite{Mantysaari:2018nng}).  As $F_2$ is sensitive to both the size and density of the target, we additionally require that the transverse size of the proton at the initial $x_0=0.01$ is compatible with the HERA measurement of the $t$ slope of the coherent $J/\Psi$ production. 

The free parameters of the model control the size and the saturation scale at the initial Bjorken-$x$, for which we take $x=0.01$. Additionally, the value of the coupling constant $\as$, or the coordinate space scale $\lqcd$ which controls where the running coupling is evaluated, is taken as a free parameter. Additionally, long distance Coulomb tails are regulated. For more details, see Ref.~\cite{Mantysaari:2018zdd}. 

\begin{figure*}[tb]
\centering
\begin{minipage}{0.48\textwidth}
		\includegraphics[width=\textwidth]{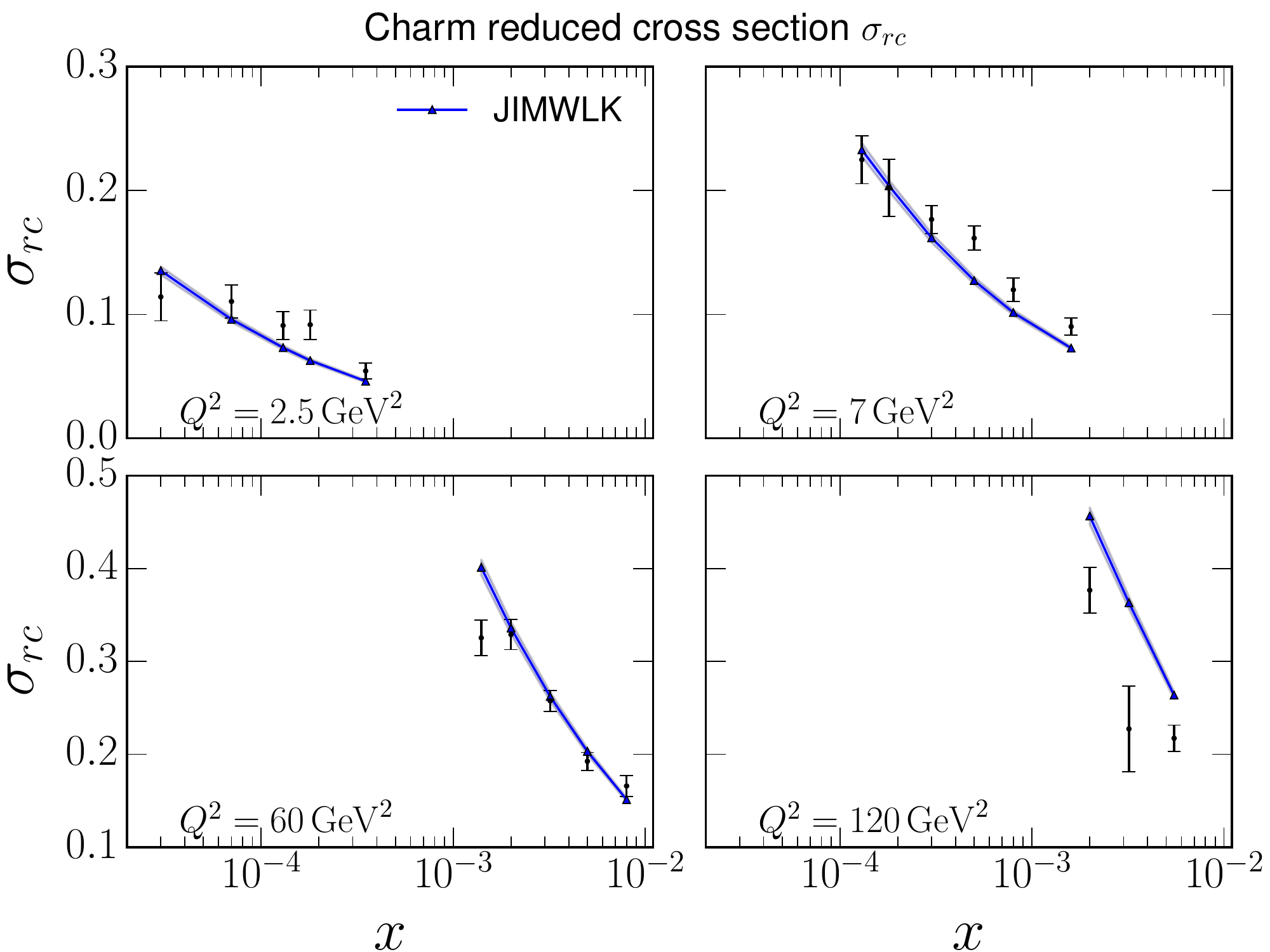} 
				\caption{Description of the HERA charm production data~\cite{H1:2018flt}. Figure from Ref.~\cite{Mantysaari:2018zdd}.		}
		\label{fig:sigmar}
\end{minipage}
\quad
\begin{minipage}{0.48\textwidth}
\centering
		\includegraphics[width=\textwidth]{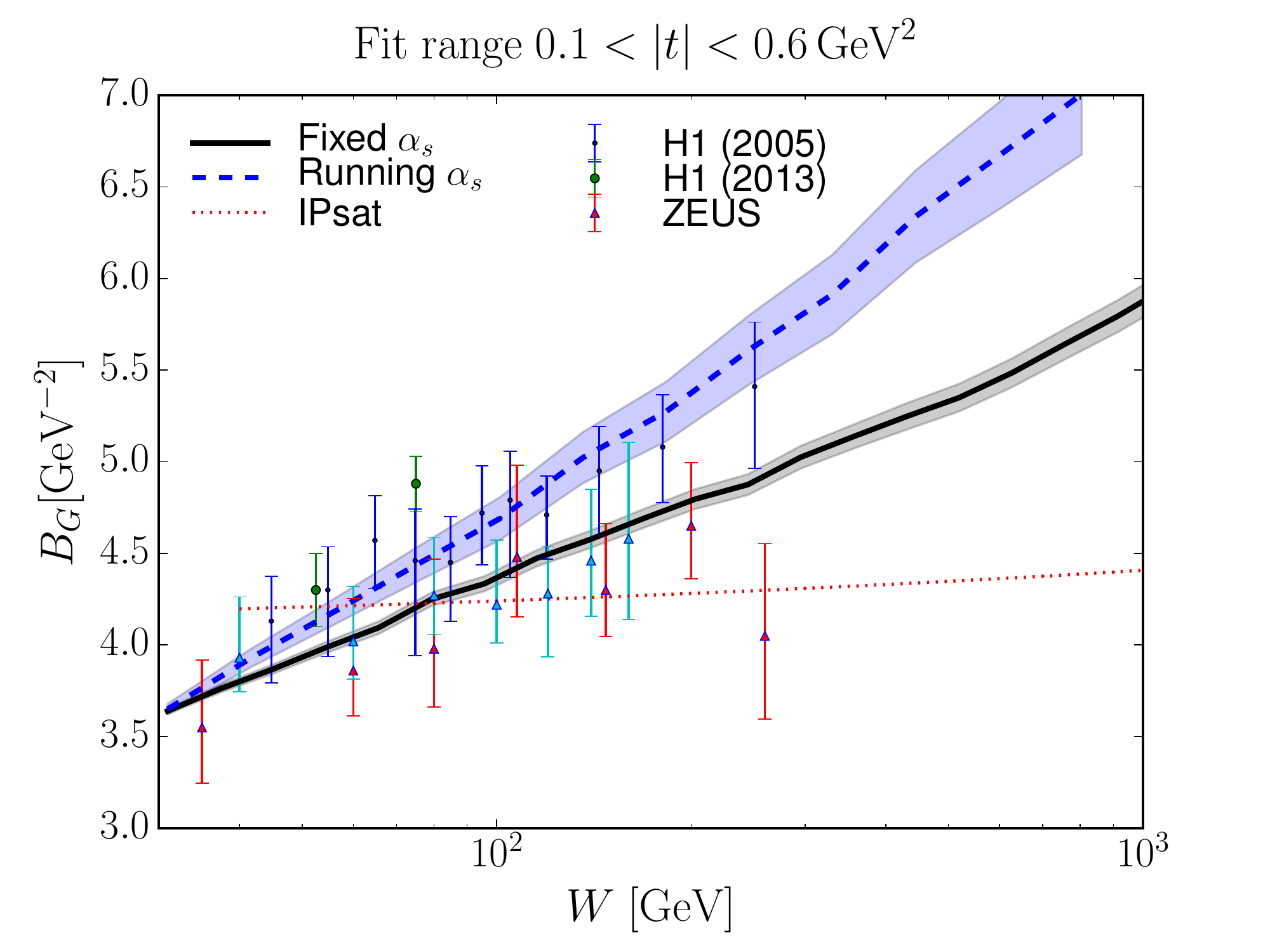} 
				\caption{Slope of the diffractive $J/\Psi$ production compared with the HERA data~\cite{Chekanov:2002xi,Aktas:2005xu,Alexa:2013xxa}. Figure from Ref.~\cite{Mantysaari:2018zdd}.}
				
		\label{fig:slope_mdep_mint_0.1}
\end{minipage}
\end{figure*}

The description of the charm reduced cross section data is shown in Fig.~\ref{fig:sigmar}. In the figure the results with fixed coupling JIWMLK evolution are shown, the result at running coupling being identical. The $Q^2$ evolution speed is faster than in the data, which is expected as HERA data is known to prefer (at leading order) an anomalous dimension $\gamma > 1$ unlike $\gamma=1$ in the MV model, see e.g.~\cite{Lappi:2013zma}. As discussed in more detail in Ref.~\cite{Mantysaari:2018zdd}, a similar effect can be introduced in this framework by filtering out high-frequency modes.

When the dipole size becomes comparable to the size of the proton target and the quarks miss the target, the dipole scattering amplitude drops to zero. Consequently, the contribution from large dipoles (compared to the proton size) is heavily suppressed in our framework. As shown in  Ref.~\cite{Mantysaari:2018zdd}, in order to describe inclusive structure functions a large non-perturbative contribution needs to be included. Additionally, this also causes our prediction for the absolute $J/\Psi$ production cross section to underestimate the data.

The evolution of the proton size is shown in Fig.~\ref{fig:slope_mdep_mint_0.1}. With running coupling the proton is found to grow faster than at fixed coupling, even though in both cases the resulting charm cross sections are practically identical. This can be understood, as at running coupling the evolution of the short-wavelength modes is suppressed relative to long wavelength ones that dominate the proton size measure at low $|t|$.

\begin{figure*}[tb]
\centering
\begin{minipage}{0.48\textwidth}
		\includegraphics[width=\textwidth]{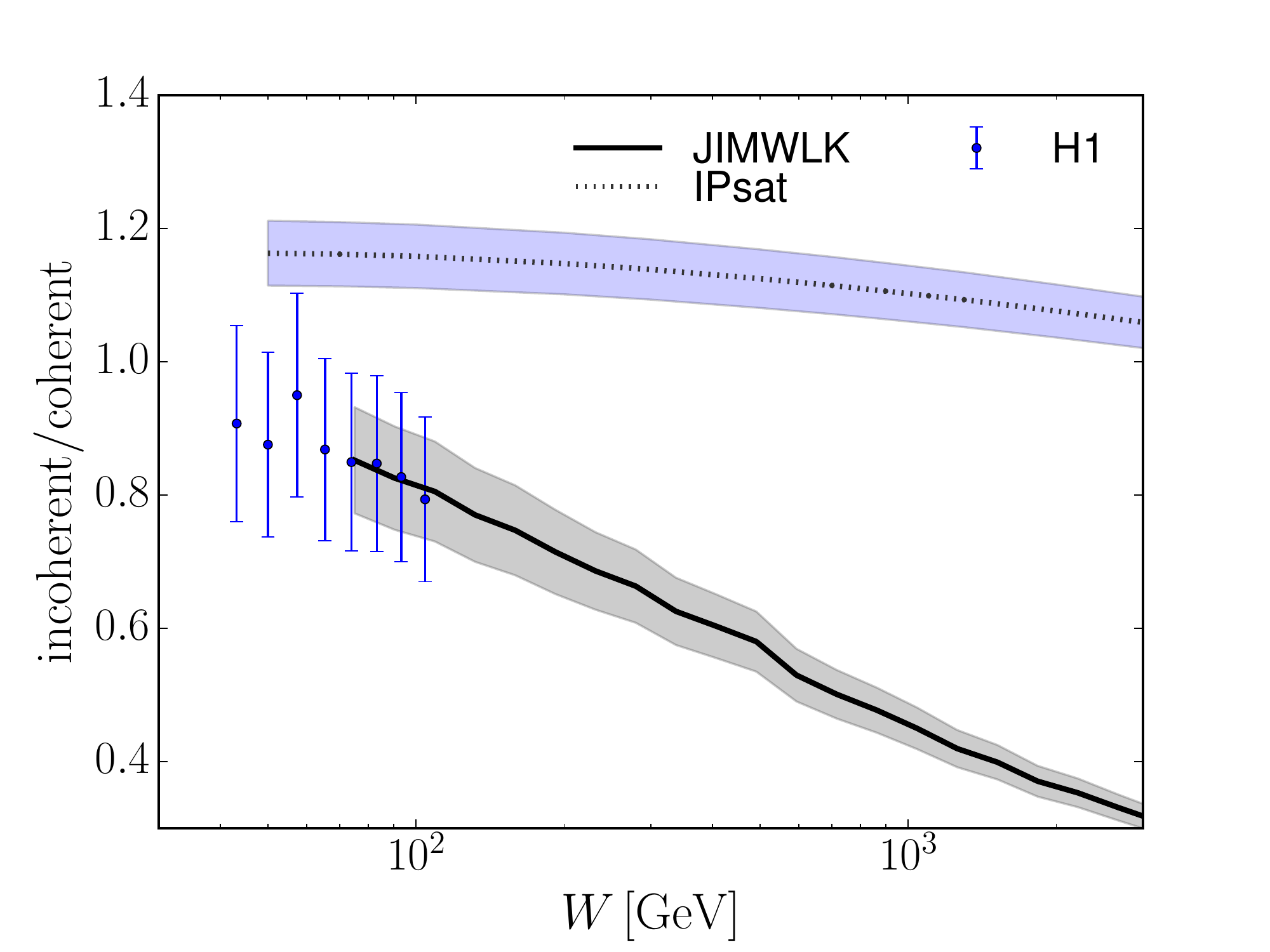} 
				\caption{Ratio of incoherent to coherent cross section. The experimental uncertainties are computed by assuming independent errors for total coherent and incoherent cross sections~\cite{Alexa:2013xxa}.	}
		\label{fig:incohcohratio}
\end{minipage}
\quad
\begin{minipage}{0.48\textwidth}
\centering
		\includegraphics[width=\textwidth]{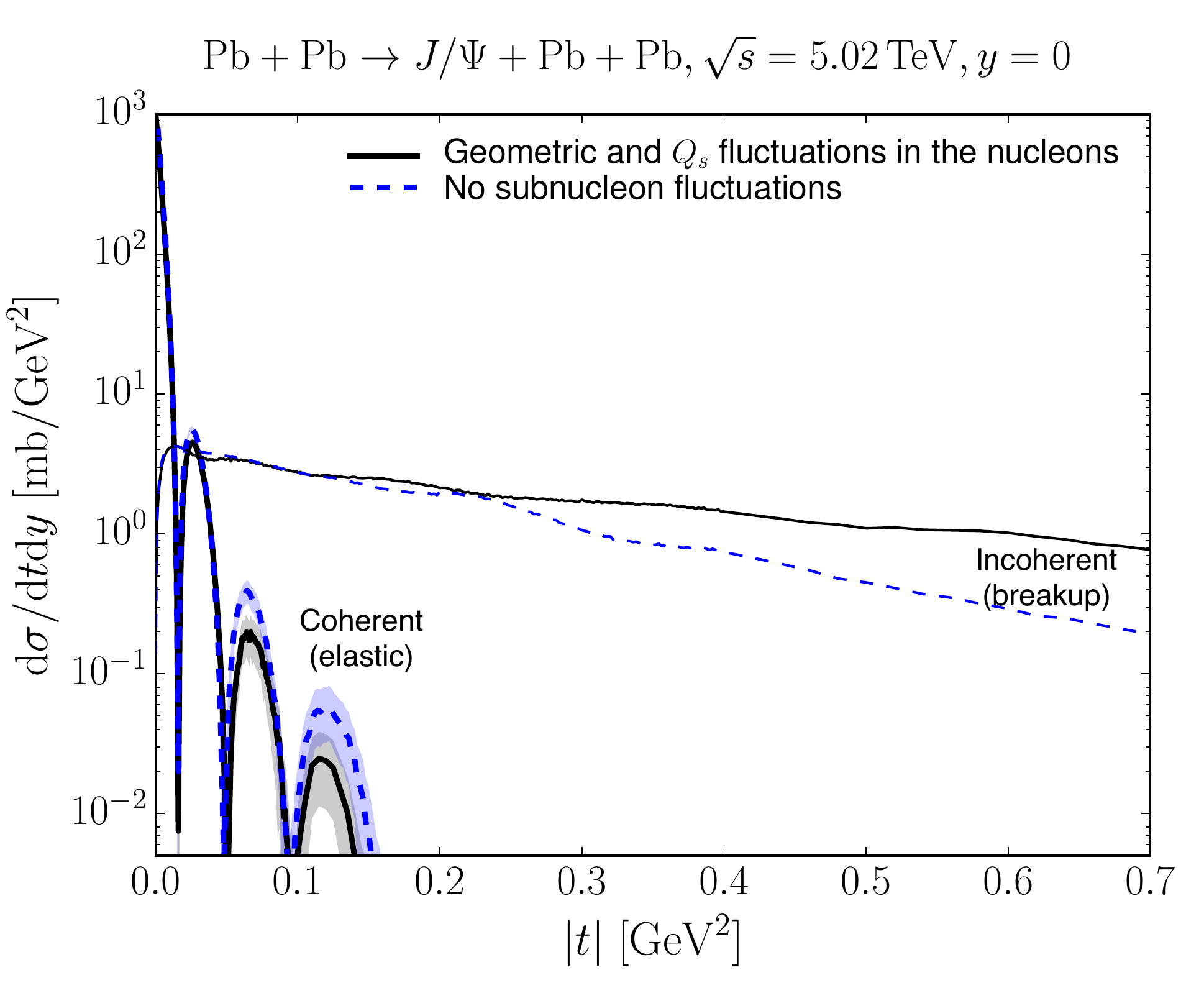} 
				\caption{Coherent and incoherent $J/\Psi$ production off a heavy nucleus with and without nucleon shape fluctuations from Ref.~\cite{Mantysaari:2017dwh}.}
				
		\label{fig:gamma-pb-t}
\end{minipage}
\end{figure*}

To study the evolution of the proton geometry, we fix the parametrization that describes the fluctuating geometry at $W=75\gev$ (corresponding to $x\approx 10^{-3}$) by comparing with the HERA $J/\Psi$ data as in Ref.~\cite{Mantysaari:2016ykx}. Then, we use the same JIMWLK evolution than previously for which the strong coupling constant is constrained by the charm production data, and compute the incoherent to coherent cross section ratio. The results are shown in Fig.~\ref{fig:incohcohratio} where a good agreement with the H1 data is found. For comparison, in Fig.~\ref{fig:incohcohratio} the result from an IPsat model calculation, where there is no geometry evolution, is shown. In that case the proton does not get less lumpy at small $x$, and the cross section ratio is practically flat.

To study the effect of the nucleon shape fluctuations in nuclei, we finally calculate $J/\Psi$ production in ultraperipheral lead-lead collisions at the LHC. The results with and without nucleon shape fluctuations are shown in Fig.~\ref{fig:gamma-pb-t}. In case of the coherent cross section, the fluctuations have no effect as the average density profile of the nucleus is not significantly modified. In case of incoherent $J/\Psi$ production, when $|t|$ is large enough that one becomes sensitive to distance scales comparable to the hot spots that make up the nucleons, the nucleon shape fluctuations enhance the incoherent cross section significantly. See Ref.~\cite{Mantysaari:2017dwh} for details.

\section{Conclusions}

We have fitted the initial condition to the JIMWLK evolution to the HERA charm production data. The resulting evolution of the proton structure is compatible with the measurement of the (gluonic) size of the proton measured in exclusive $J/\Psi$ production, but inclusive structure function and $J/\Psi$ production cross section are underestimated. The JIMWLK evolution makes it possible to calculate the Bjorken-$x$ evolution of the event-by-event fluctuating proton structure. We find that the evolution washes out the initial hot spot structure, and the resulting  cross section ratio and its energy dependence is compatible with the H1 data.

Our calculation of the exclusive $J/\Psi$ production off lead in ultraperipheral heavy ion collisions at the LHC shows that the nucleon shape fluctuations have a significant effect on the incoherent vector meson production starting at moderate $|t|\sim 0.2\gev^2$. Measuring the incoherent $J/\Psi$ spectra at the LHC will then directly test if the nucleon shape fluctuations are visible within the heavy nucleus.

\subsection*{Acknowledgments}
H. M. is supported by European Research Council, Grant ERC-2015-CoG-681707. BPS acknowledges a DOE Office of Science Early Career Award. Computing resources of the National Energy Research Scientific Computing Center (supported by the DOE Contract No. DE-AC02-05CH11231), and of the CSC – IT Center for Science in Espoo, Finland, were used in this work.





\bibliographystyle{elsarticle-num}
\bibliography{../../refs.bib}







\end{document}